# Adjoint-optimized nanoscale light extractor for nitrogen-vacancy centers in diamond

Raymond A. Wambold[1], Zhaoning Yu[1,2], Yuzhe Xiao[1], Benjamin Bachman[3], Gabriel Jaffe[2], Shimon Kolkowitz[2], Jennifer T. Choy[4,1,5], Mark A. Eriksson[2], Robert J. Hamers[3], Mikhail A. Kats[1,2,5]

1. Department of Electrical and Computer Engineering, University of Wisconsin – Madison
2. Department of Physics, University of Wisconsin – Madison
3. Department of Chemistry, University of Wisconsin – Madison
4. Department of Engineering Physics, University of Wisconsin – Madison
5. Department of Materials Science and Engineering, University of Wisconsin - Madison

**Abstract**

We designed a nanoscale light extractor (NLE) for the efficient outcoupling and beaming of broadband light emitted by shallow, negatively charged nitrogen-vacancy (NV) centers in bulk diamond. The NLE consists of a patterned silicon layer on diamond and requires no etching of the diamond surface. Our design process is based on adjoint optimization using broadband time-domain simulations and yields structures that are inherently robust to positioning and fabrication errors. Our NLE functions like a transmission antenna for the NV center, enhancing the optical power extracted from an NV center positioned 10 nm below the diamond surface by a factor of more than 35, and beaming the light into a ±30° cone in the far field. This approach to light extraction can be readily adapted to other solid-state color centers.

**Introduction**

Negatively charged nitrogen-vacancy (NV) centers in diamond are optical emitters whose level-structure is highly sensitive to external perturbations, which makes them excellent sensors of highly localized electric and magnetic fields, temperature, and strain [1]–[5]. NV centers are of great interest for quantum computing and communication [6]–[10] and the study of quantum phenomena such as quantum entanglement and superposition [11], [12]. However, efficiently extracting NV fluorescence is often challenging due to the high index of refraction in diamond (~2.4), which results in high reflectance at the diamond-air interfaces and total internal reflection for emission angles larger than the critical angle. Previous attempts to extract more light from bulk diamond primarily involved the etching of the diamond itself (a complicated fabrication process that can adversely affect NV properties such as spin coherence) [13]–[19] or fabricating structures that still required a high numerical aperture oil-immersion objective to efficiently collect the emission (which adds system complexity and is detrimental to sensing applications) [20]–[23]. Furthermore, precision etching of diamond around NV centers can be a substantial challenge and can damage the surface of diamond, resulting in roughness and modifying the chemical termination [24], which can degrade the quantum properties of NV centers [25], [26].

Here, we design a silicon-based nanoscale light extractor (NLE) that sits on the top of a flat, unpatterned diamond surface and can enhance the optical output of near-surface NV emitters by more than 35× compared to the unpatterned case, directing the light into a narrow cone that can be easily collected with low-NA optical systems. Our NLE consists of a patterned silicon structure on top of the diamond surface (Fig. 1), directly above a shallow NV center (< 300 nm below the surface). The proximity of a resonant high-index dielectric structure close to the emitter enables near-field coupling to and broadband Purcell enhancement of the emitter [23], [27]. We designed the silicon NLE using an adjoint-optimization method, optimizing for NV emission funneled into a narrow cone into the far field. This approach both increases collection efficiency and enhances the radiative emission rate of the NV center.

Our approach focuses on broadband emission enhancement of negatively charged NV centers close to the diamond surface, which is especially useful for sensing applications in which measurement sensitivity is limited by the number of collected photons across the entire NV emission spectrum.

**Figure of merit for broadband NV emission extraction**

We targeted our design towards negatively charged NV centers in [100] diamond. The NV axis is at an angle of 54.7° with respect to the normal to the diamond surface in this orientation [28], and the optical dipole moments are orthogonal to the NV axis [29]. At room temperature, the emission from an NV center is expected to be unpolarized [30] and can therefore be simulated by the incoherent sum of emitted intensities from any two orthogonal linear dipoles that are also orthogonal to the NV axis. In bulk diamond, negatively charged NV centers have optical transitions at ~637 nm (the location of the zero-phonon line in the spectrum), but fluoresce over a bandwidth of >150 nm due to vibrational side bands in the diamond [28], [31]. Therefore, we define a spectrum-averaged figure of merit ($FoM$) to quantify the degree of light extraction:

$$FoM = \frac{\int I_{NV}(\lambda) \times \eta(\lambda)\, d\lambda}{\int I_{NV}(\lambda)\, d\lambda} \qquad (1)$$

Here, $I_{NV}(\lambda)$ is the normalized emission spectrum of the NV centers in bulk diamond taken from ref. [32] and $\eta(\lambda)$ is the extraction efficiency of our NLE, which is defined as the number of photons emitted into free-space in the presence of the NLE divided by the number of photons emitted into free-space by that same NV (i.e., same depth and orientation) with no NLE present, at a wavelength $\lambda$. Note that $\eta(\lambda)$ includes both collection and Purcell enhancements. For example, an NLE that results in 10 times as many photons emitted into free space at every wavelength would have $\eta(\lambda) = 10$ and $FoM = 10$.

For all plots in this paper, the bounds of the integral in Eqn. (1) are set to 635 – 800 nm to cover the NV zero-phonon line and phonon sideband emission spectrum. Our NLE achieves broadband NV fluorescence enhancement, which is particularly useful for sensing applications using shallow NVs, but the design approach can readily be adapted to focus on more-narrow spectral ranges (e.g., the zero-phonon line alone).

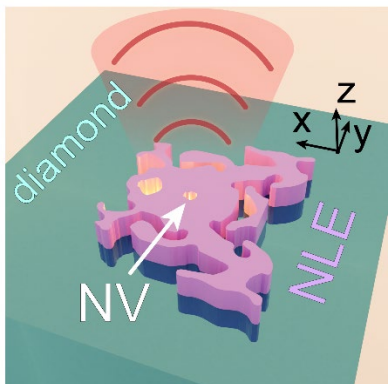

Figure 1. Schematic of the nanoscale light extractor (NLE), which sits on the diamond surface above an NV center and directs fluorescence out of the diamond and into a narrow cone in the far field. The schematic is a render of the actual optimized structure reported below.

**Adjoint-optimization method**

To design a high-performance NLE, we used adjoint optimization—a design technique that has been used extensively in mechanical engineering and has recently been applied to the design of optical metasurfaces and other photonic structures [33]–[35]. For free-form structures that are allowed to evolve in three dimensions, the optimization process consists of the evolution of the structure, defined by a refractive-index profile $n(r,\lambda)$, towards maximizing the overlap between a forward simulation field $[E_{\text{fwd}}(r,\lambda)]$ and a specific adjoint simulation field $[E_{\text{adj}}(r,\lambda)]$, where $r$ is a position within the structure.

Due to the incoherent and unpolarized emission pattern of the NV center at room temperature, our approach was to maximize extraction from two orthogonal optical-dipole orientations [29]. We used two pairs of forward/adjoint simulations for these two orientations. The first of our forward simulations is sourced by an electric dipole 10 nm below the diamond surface, with the dipole moment oriented in the XZ plane and tilted at an angle of 35.3° off the Z axis. The second was sourced by a dipole placed at the same position and oriented parallel to the Y axis. Note that NV centers in [100] diamond can be oriented in one of four directions along the ⟨111⟩ crystal axes. Thus to maximize performance, prior characterization can be done to align the NLE in the proper direction during fabrication [39]. Two adjoint Gaussian beams were used with orthogonal polarizations; one polarized along the X axis and the other polarized along the Y axis to pair with the forward dipoles in the XZ plane and along the Y axis, respectively.

We used finite-difference time-domain (FDTD) simulations (implemented in Lumerical FDTD [36]) to determine the forward and adjoint fields in the optimization region across the spectrum of interest for both optimization pairs. At each optimization generation, we calculated a figure-of-merit gradient similar to that of ref. [37], but averaged across the forward/adjoint simulation pairs:

$$G(r,\lambda) \propto \langle n(r,\lambda) \cdot Re[E_{\text{fwd}}(r,\lambda) \cdot E_{adj}(r,\lambda)] \rangle_{pairs} \qquad (2)$$

For the case of only one forward/adjoint pair, a positive value of $G(r,\lambda)$ indicates that a small increase in the refractive index $n(r,\lambda)$ at position $r$ will result in a stronger overlap between the forward and adjoint fields. For more than one pair, like here, $G(r,\lambda)$ is a compromise quantity balancing the performance between all pairs. This causes the optimization algorithm to move in the direction that maximizes the gradient with respect to both pairs, and, while not necessarily reaching the same performance levels as with a single pair, is required in our case due to the nature of the NV emission. Furthermore, because it is typically not possible to separately engineer the index at different wavelengths, we define a wavelength-weighted gradient:

$$G(r) = \frac{1}{\Delta\lambda} \int I_{NV}(\lambda) G(r,\lambda) d\lambda \qquad (3)$$

To design a structure that can realistically be fabricated using top-down techniques, there should ideally be no material variance in the vertical (Z) direction. We impose this constraint in our optimization by 1) forcing the index in a single column along Z to be constant, and 2) averaging $G(r)$ along each column during the update step such that a single $G$ is applied to each column given by:

$$G(\bar{r}) = \int_{z_{min}}^{z_{max}} \frac{G(r)}{z_{max}-z_{min}} dz \qquad (4)$$

(similar to, e.g., ref. [37]), where $\bar{r}$ is the two-dimensional position vector in the $(x, y)$ plane.

Figure 2(A) shows the design process for our NLE. To start, the optimization is seeded with a random continuous distribution of complex refractive index, taking values between that of air ($n_{air}$ =1) and crystalline silicon ($n_{Si}(\lambda)$, from ref. [38]), calculated by $n(\bar{r}, \lambda) = p(\bar{r}) \cdot n_{Si}(\lambda) + [1 - p(\bar{r})] \cdot n_{air}$ where $p(\bar{r})$ ranges between 0 and 1, such that if $p = 1$, then $n = n_{Si}$. $G(\bar{r})$ is then used to update the position-dependent refractive-index profile according to $p_{new}(\bar{r}) = p_{old}(\bar{r}) + c \cdot G(\bar{r})$, where $c$ is a normalization factor. In principle, this process can be iterated (while evolving $c$), until $p(\bar{r})$ converges to yield an optimized index profile.

However, the optimization method described above yields an optimized profile $p(\bar{r})$ corresponding to a continuum of index values $n(\bar{r}, \lambda)$. To evolve the index distribution into a binary structure of silicon and air that can be made by lithography and etching, and also ensuring that the feature sizes are not too small, we followed the methods of Sigmund in ref. [40]. This approach applies a conically shaped blurring function to $p(\bar{r})$ at each iteration to smooth the index distribution to remove features that are smaller than the cone radius, $R$. For our optimization we used a conical blurring function with $R = 40$ nm to result in structures that can be readily made with most electron-beam lithography systems. The new blurred profile $p_{blur}(\bar{r})$, is created according to [40]:

$$p_{blur}(\bar{r}) = \int \frac{w(\bar{r}, \bar{r}') \cdot p(\bar{r})}{w(\bar{r}, \bar{r}')} d\bar{r}' \qquad (5)$$

$$w(\bar{r}, \bar{r}') = \begin{cases} 0, & \|\bar{r}' - \bar{r}\| > R \\ R - \|\bar{r}' - \bar{r}\|, & \|\bar{r}' - \bar{r}\| \le R \end{cases} \qquad (6)$$

where $w(\bar{r}, \bar{r}')$ is a weight function that becomes larger closer to position $\bar{r}$.

In addition to blurring, a binary push function is required to finish the optimization with a fully binary structure (i.e., in our case, at each point in the optimization region the final material should be either air or silicon). The binarization method found in ref. [40] and [35] was the guide for our implementation. Specifically, this method modifies the blurred element matrix as follows:

$$p_{bin}(\bar{r}) = \begin{cases} \alpha e^{-\frac{\beta(\alpha - p_{blur}(\bar{r}))}{\alpha}} - (\alpha - p_{blur}(\bar{r}))e^{-\beta}, & 0 \le p(\bar{r}) \le \alpha \\ 1 - (1-\alpha)e^{-\frac{\beta(p_{blur}(\bar{r}) - \alpha)}{1-\alpha}} - (\alpha - p_{blur}(\bar{r}))e^{-\beta}, & \alpha < p(\bar{r}) \le 1 \end{cases} \qquad (7)$$

where $p_{bin}(\bar{r})$ is the modified (new) value of $p$ at position $\bar{r}$. Here, $\alpha$ is a cutoff parameter that allows us to define dilated and eroded edges (to account for fabrication errors), but is set to the midpoint value of 0.5 during the optimization process. The parameter $\beta$ controls the strength of binarization applied to the structure. A $\beta$-value of 0 gives no increased binarization to the structure leading to $p_{bin} = p$. As $\beta$ increases, values of $p < \alpha$ are pushed towards zero while values of $p > \alpha$ are pushed towards one. Like the blurring function, the binarization function is applied at every step of the optimization process, with $\beta$ starting at 0 and increasing, until a binary structure is achieved at the end of the optimization [~ iteration 400 in Fig. 2(C)]. We found that to further remove small features and move the structure away from sharp local maxima, it was beneficial to apply a second blurring function to $p(\bar{r})$ once every 35 iterations late in the optimization process (after iteration 305 in Fig. 2(C)).

We leveraged the inherently broadband nature of the FDTD method to simulate and optimize the performance of our device across the entire NV emission spectrum, simultaneously achieving a design that is robust to errors in fabrication. Previous works using adjoint optimization with a frequency-domain electromagnetic solver achieve fabrication robustness through simulating dilated and eroded devices each iteration, which corresponds to a three-fold increase in the number of simulations required [35], [41]–[43]. In our optimizations, fabrication robustness was built-in automatically due our requirement for broadband performance together with the scale invariance of Maxwell's equations (i.e., larger structures at longer wavelength behave similarly to smaller structures at shorter wavelength). This correspondence between broadband performance and shape robustness has been observed previously for other inverse-design approaches [44], [45].

Note that our adjoint optimization setup that uses $G(\bar{r})$ to evolve the structure does not directly maximize the $FoM$ in Eqn. (1), because the formalism described above requires the selection of specific optical waveforms for the forward and adjoint fields, whereas the our $FoM$ does not depend on the particular spatial mode(s) in which light escapes to free space, or the phase of that mode relative to the dipole source. Here, we selected our adjoint source to be a Gaussian beam with a diffraction angle of ±30° injected toward the diamond surface along the Z axis. To relax the phase degree of freedom, we performed multiple optimization runs in which we varied the relative phases of the forward and adjoint sources from 0 to $2\pi$ in steps of $\pi/2$. Due to having two forward and two adjoint sources, this led to 16 total combinations of phases for a given seed. To reduce computation time, only the best 4 phase combinations were selected to finish after 30 generations. While not pursued here, the same type of approach can be taken for accounting for the polarization of the sources.

**Designing the nanoscale light extractor**

To find the ideal height, we first ran a series of a series of 2D optimizations, sweeping over a range of heights (Fig. 2A,B). Like in the later 3D simulations, the 2D simulations were set up with the dipole embedded 10 nm into diamond, and we used two orthogonal dipole orientations in a plane perpendicular to the NV axis. We swept the height from 50 to 600 nm in steps of 50 nm and ran 5 optimization runs for each height. Figure 2(B) shows the average of our $FoM$ for each height, with the oscillations reminiscent of Fabry-Perot fringes. Based on these results, we decided to run the full 3D optimization of the NLE for a height of 300 nm. This thickness is a common device-layer thickness in silicon-on-insulator (SOI) technology [46] and we note that SOI lends itself well to fabricating this structure as crystalline silicon device layers can be transferred to diamond using membrane-transfer techniques [47].

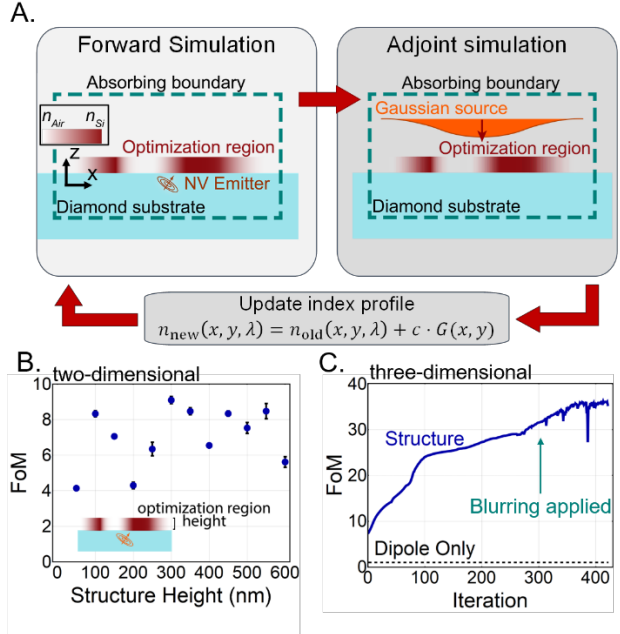

Figure 2. **A.** Visualization of the optimization routine, which evolves an index profile situated above an NV emitter. In the forward simulation, the light source is a dipole at the location of the NV. The sources in the adjoint simulations are two orthogonally polarized Gaussian beams injected from free space toward the structure. The sensitivity gradient, $G(\bar{r}) = G(x,y)$, is calculated and then used to evolve the index profile. The updated profile is then used in the next iteration. **B.** A sweep of the structure height (i.e., the thickness of the Si membrane), running five full 2D optimization cycles for each height. We found a height of approximately 300 nm to be optimal over the range of the sweep. The error bars represent the variance at each height. The variance is very small for some heights, so the error bars are not visible. **C.** FoM vs. the iteration number for a full 3D optimization run to generate our NLE. The dips are due to the application of the secondary blurring later in the optimization.

The full 3D optimization run can be seen in Fig. 2(C). The $FoM$ increases as the optimization progresses, with occasional dips due to the implementation of the secondary blurring function. The optimization terminates when the device is sufficiently binarized, such that the structure is entirely comprised of air and silicon, and the FoM is no longer improving significantly.

**Results**

The structure of our optimized NLE is shown in Fig. 3(A), with the extraction efficiency in Fig. 3(B). Though the NLE was optimized for an NV depth of 10 nm, we also show the results for depths of 5 and 15 nm but with the same NLE structure. The extraction efficiency increases for NVs closer to the structure due to increased coupling with the NLE, which leads to larger Purcell enhancement [48]. For an NV depth of 10 nm, the Purcell enhancement of our device is ~3, averaged across the emission spectrum. Here, we calculate the Purcell enhancement by dividing the power emitted by the dipole source in the presence of the interface and the NLE by the power emitted by the dipole in homogeneous diamond (no interface).

Figure 3(C, D) shows the near- and far-field radiation patterns. The NLE can shape the emitted fields from the dipole into a beam that is approximately Gaussian. The beam angle is slightly offset from the normal (~7.5°) but the bulk of the beamed power fits within a +/- 30° cone. The fields in Fig. 3(C,D) are for a wavelength of 675 nm, but the beaming persists across the NV spectrum and the peak emission angle does not deviate by more than $\pm 5°$ (see supplemental for details). For an objective with numerical aperture of 0.75, the collection efficiency, defined as the fraction of emitted optical power that can be collected, is around 40% in the 635 – 670 nm range, and above 25% across the entire 635 – 800 nm range (supplemental S4).

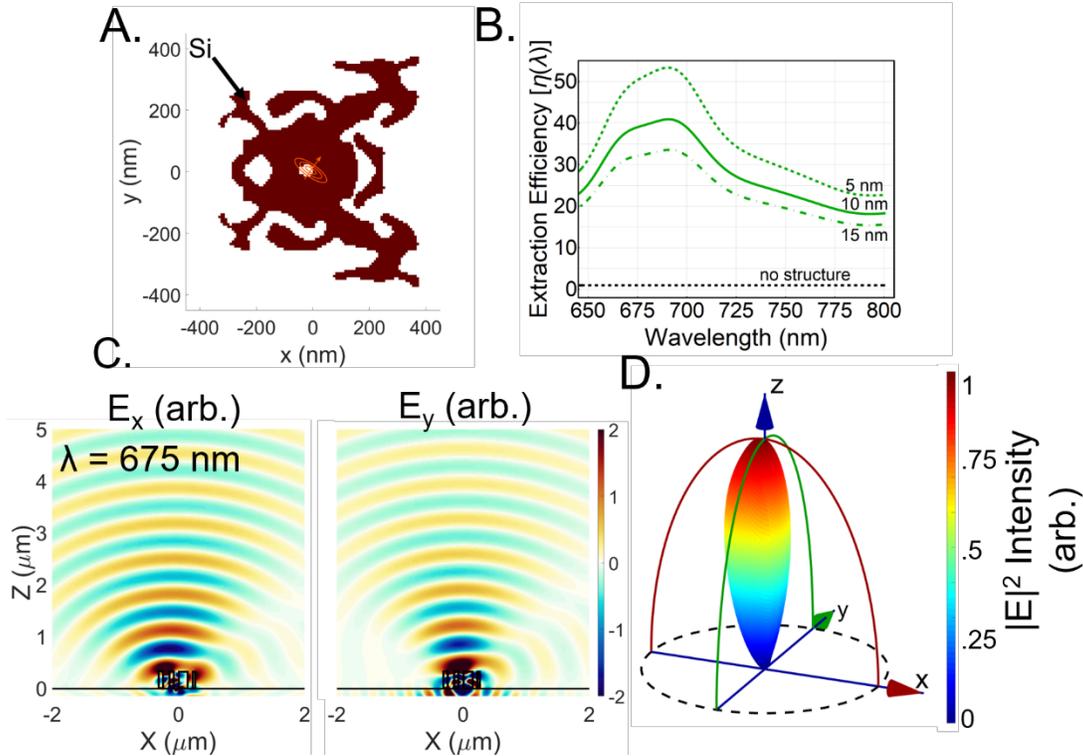

Figure 3. **A.** Top-down view of the final NLE optimized for an NV depth of 10 nm. **B.** Extraction efficiency $[\eta(\lambda)]$ of the NLE for NV depths of 5, 10, and 15 nm below the diamond/air interface. **C,D.** Snapshot of the near and meso- fields of the emitted electric field from (left) a dipole in the X-Z plane, and (right) a dipole in the Y direction. **D.** Intensity far-field averaged over the two dipole orientations. In C,D, the plots are at a wavelength of 675 nm. The bulk of the beamed power fits within a 60° cone in the far field.

Due to the broadband spectrum incorporated into the optimization, our devices display considerable tolerance to various fabrication defects as well as robustness to alignment errors (Fig. 4). Though our optimized NLE structure was based on a fixed NV depth of 10 nm, we also simulated its performance for a variety of depths [Fig. 4(A)]. The NLE performs better as the dipole gets closer to the surface due to enhanced near-field coupling. The performance falls off by a factor of 2 at a depth of 40 nm, yet still maintains enhancements of about 15 times the emission of an NV with no NLE. Even down to a dipole depth of 300 nm, the NLE is able to increase the output of the NV by a factor of 3. Note that for NVs at depths substantially different than 10 nm, a more-effective design can very likely be found using the optimization method described above.

The NLE shows minimal performance loss from errors in rotational alignment in the range of -20° to +20° [Fig. 4(B)]. The full 360° rotational plot can be found in the supplemental. We also tested the case where the NLE was offset by some amount from the central dipole position [Fig. 4(D)], and found that the *FoM* remains above 25 for X offsets of +/-30 nm and Y offsets of +/- 40 nm.

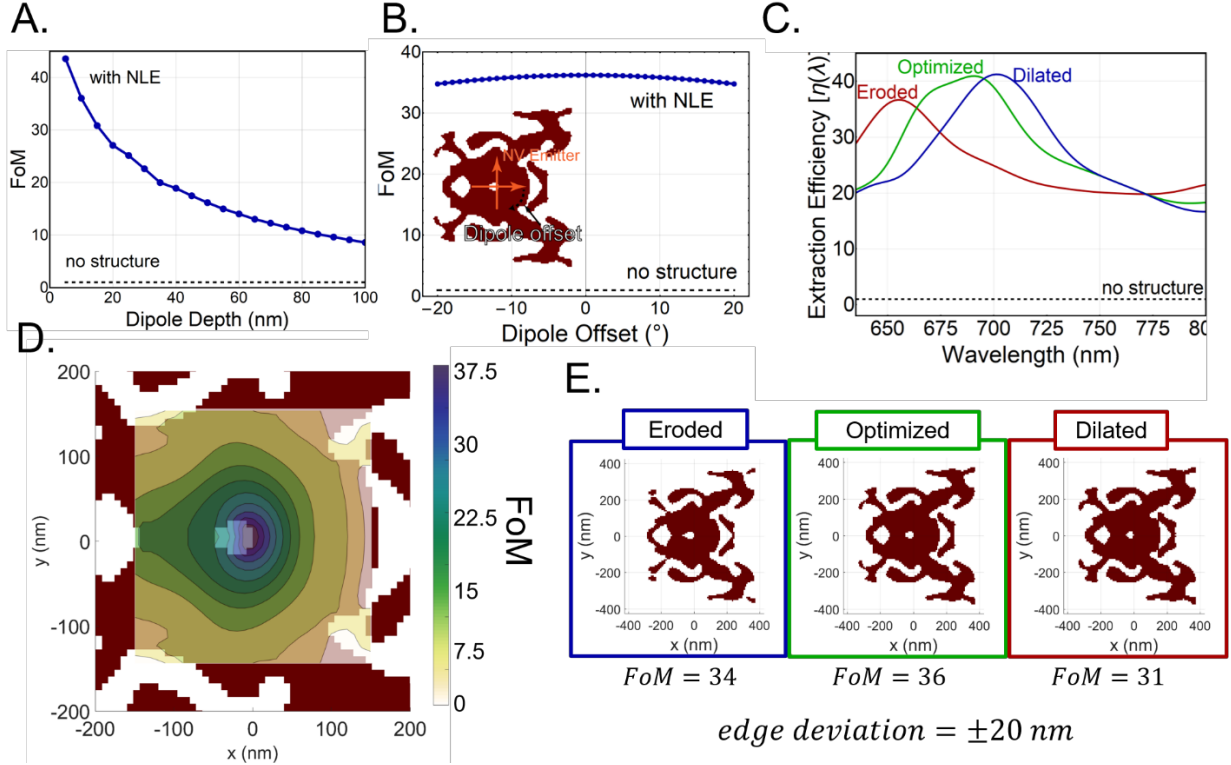

Figure 4. **A.** The NLE maintains good performance through a range of depths, with increasing FoM for NVs closer to the surface. **B.** FoM dependence on the NV emitter angle emulating angular alignment errors of the NLE. **C.** Demonstration of the fabrication robustness of the optimized device for an NV depth of 10 nm. Eroded and dilated structures are based on the optimized structure with edge deviation of +/- 20 nm to represent fabrication under/over-etching, respectively. Due to the broadband nature of our optimization, the NLE shows strong tolerance to fabrication errors. **D.** Tolerance of the NLE to lateral offsets of the NV center. The FoM remains above 25 for X offsets of +/- 30 nm and Y offsets of +/- 40 nm. **e.** The geometries of the eroded, optimized, and dilated devices simulated in c. The edge deviation refers to how far the edges shifted inward for the eroded or outward for the dilated cases.

In practice, one can reasonably expect the fabrication process to cause deviations in edge locations, e.g., due to proximity effects in lithography [49]. Figure 4(C) shows the extraction efficiency of our eroded, optimized, and dilated structures with their index profiles shown in Fig. 4(E). We calculated the deviated structures by applying a gaussian blurring filter across the optimized pattern, and then selecting cutoff points of the blurred edges to yield new binarized structures. Here, we selected the blur and cutoff to yield edge deviations of ±20 nm. The NLE maintains good performance across the spectrum despite erosion or dilation [Fig. 2(B)].

It is instructive to compare the FoM of our NLE (~35 for NVs at a depth of 10 nm) with some existing structures in the literature designed for broadband vertical outcoupling of light from a diamond slab, with the understanding that the FoM combines the distinct collection-enhancement and Purcell mechanisms. Dielectric structures comprising etched diamond typically do not provide much Purcell enhancement and include bullseye gratings [21], vertical nanowires [50], solid-immersion metalenses [19], and parabolic reflectors [13]; we estimate that these structures have calculated FoMs of approximately 10 to 20 due entirely to collection enhancement. Resonant geometries proposed thus far have primarily relied on plasmonic enhancement in structures such as metallic cavities [14], [51] and gratings [52], with FoMs of up to ~35. The theoretical FoM of resonant metal-dielectric structures in ref. [53] can be well above 100 due to large Purcell enhancement, but the implementation would require elaborate fabrication including filling etched diamond apertures with metal, and the small mode volume limits the positioning tolerance of the NV center to within ~5 nm of the field maximum. Finally, note that unlike all of the aforementioned designs, our NLE does not require etching of the diamond, and can be fabricated by using Si membrane transfer techniques [47], [54] or direct CVD growth of Si on a diamond substrate. We do note that the exact effects of the NLE on the NV-center spin characteristics (e.g., coherence) due to the proximity of the NV to the NLE surface and material are unknown at this time. These effects are difficult to predict and are beyond the scope of this article, warranting future experimental investigations. However, our technique is likely to be less invasive than approaches that require etching the diamond near the NV, which inevitably introduces damage and additional defects.

We note that while the present manuscript was undergoing peer review, a preprint by Chakravarthi et al describing a similar approach to extracting light for NV centers, but for quantum information applications, was posted on arXiv [55].

**Conclusion**

We presented a nanoscale light extractor (NLE) designed using adjoint-optimization methods with time-domain simulations to enhance the broadband emission of NV centers in diamond. Our design not only enhances the fluorescence of the NV centers but also demonstrates exceptional beam-shaping qualities, enabling efficient collection with a low-NA lens in free space. Even given reasonable uncertainty in NV center localization and errors in device fabrication, simulations show the NLE maintains a high extraction efficiency. Our results suggest that such robustness to positioning and fabrication errors can be automatically achieved when optimizing for a broadband figure of merit. The NLE can be fabricated with conventional electron-beam lithography techniques without etching of the diamond surface. Our approach can easily be extended to other color centers in diamond, as well as material systems where increased light extraction from defect centers is desired, such as silicon carbide or hexagonal boron nitride [56], [57].

**Acknowledgements**

This material is based upon work supported by the National Science Foundation under Grant No. CHE-1839174. RW acknowledges support through the DoD SMART program and a scholarship from the Directed Energy Professional Society. Additional contributions by JTC and SK were supported by the U.S. Department of Energy (DOE), Office of Science, Basic Energy Sciences (BES) under Award #DE-SC0020313. We thank an anonymous reviewer for identifying an error in the orientation of the NV optical axis in a previous version of the manuscript.

Supplemental information:

# Adjoint-optimized nanoscale light extractor for nitrogen-vacancy centers in diamond

Raymond A. Wambold[1], Zhaoning Yu[1,2], Yuzhe Xiao[1], Benjamin Bachman[3], Gabriel Jaffe[2], Shimon Kolkowitz[2], Jennifer T. Choy[4,1,5], Mark A. Eriksson[2], Robert J. Hamers[3], Mikhail A. Kats[1,2,5]

1. Department of Electrical and Computer Engineering, University of Wisconsin – Madison
2. Department of Physics, University of Wisconsin – Madison
3. Department of Chemistry, University of Wisconsin – Madison
4. Department of Engineering Physics, University of Wisconsin – Madison
5. Department of Materials Science and Engineering, University of Wisconsin - Madison

## S1. Near-field comparison with and without the NLE

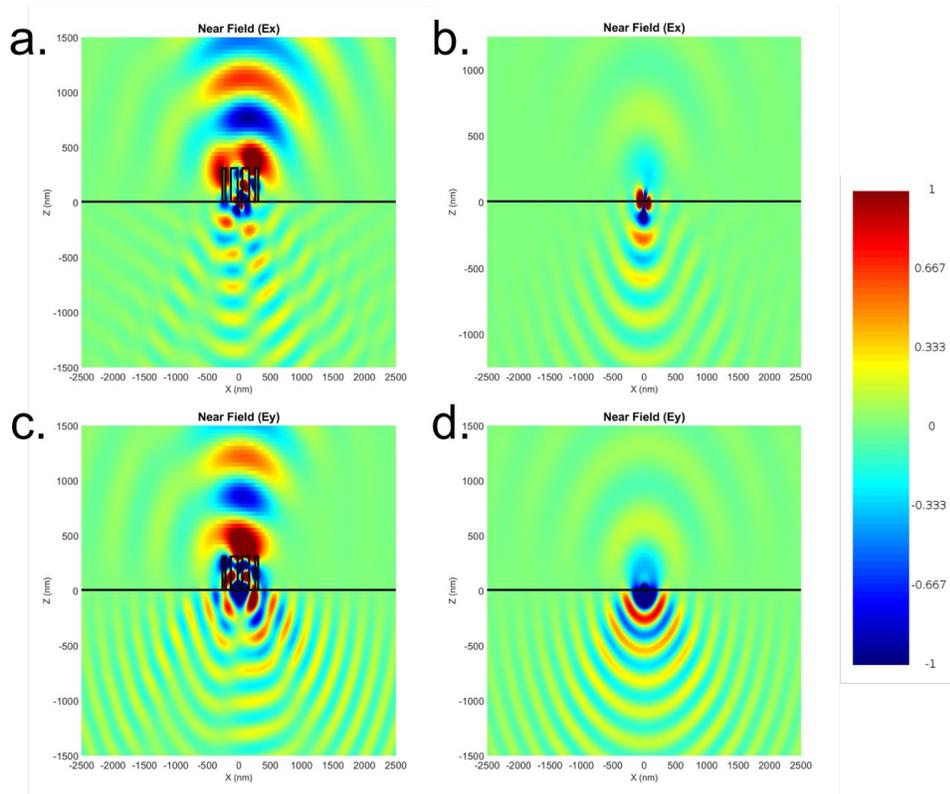

**Figure S1**. A snapshot of the electric fields in the vicinity of the NV center taken along a slice in the XZ plane at y = 0 (the center of the device and the y-coordinate of the NV), at a wavelength of 675 nm. Here we directly compare the X components of the electric field for the dipole in the XZ plane **(a)** with and **(b)** without the NLE, and the Y components of the electric field for the dipole in the XY plane **(c)** with and **(d)** without the NLE. The NLE is outlined in black for reference.

## S2. FoM vs. the NLE angle with respect to the NV orientation

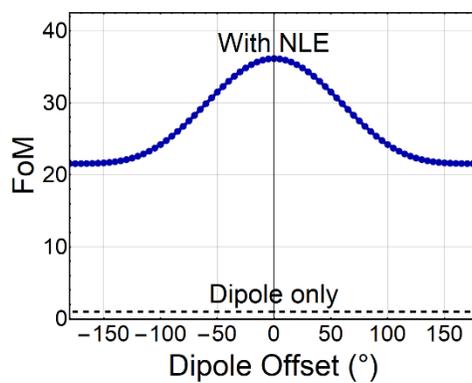

**Figure S2**. The full rotational robustness of the NLE demonstrated by rotating the NV emitter a full 360° around the Z axis.

## S3. Far-field |E|² plots for select wavelengths across 635 – 800 nm

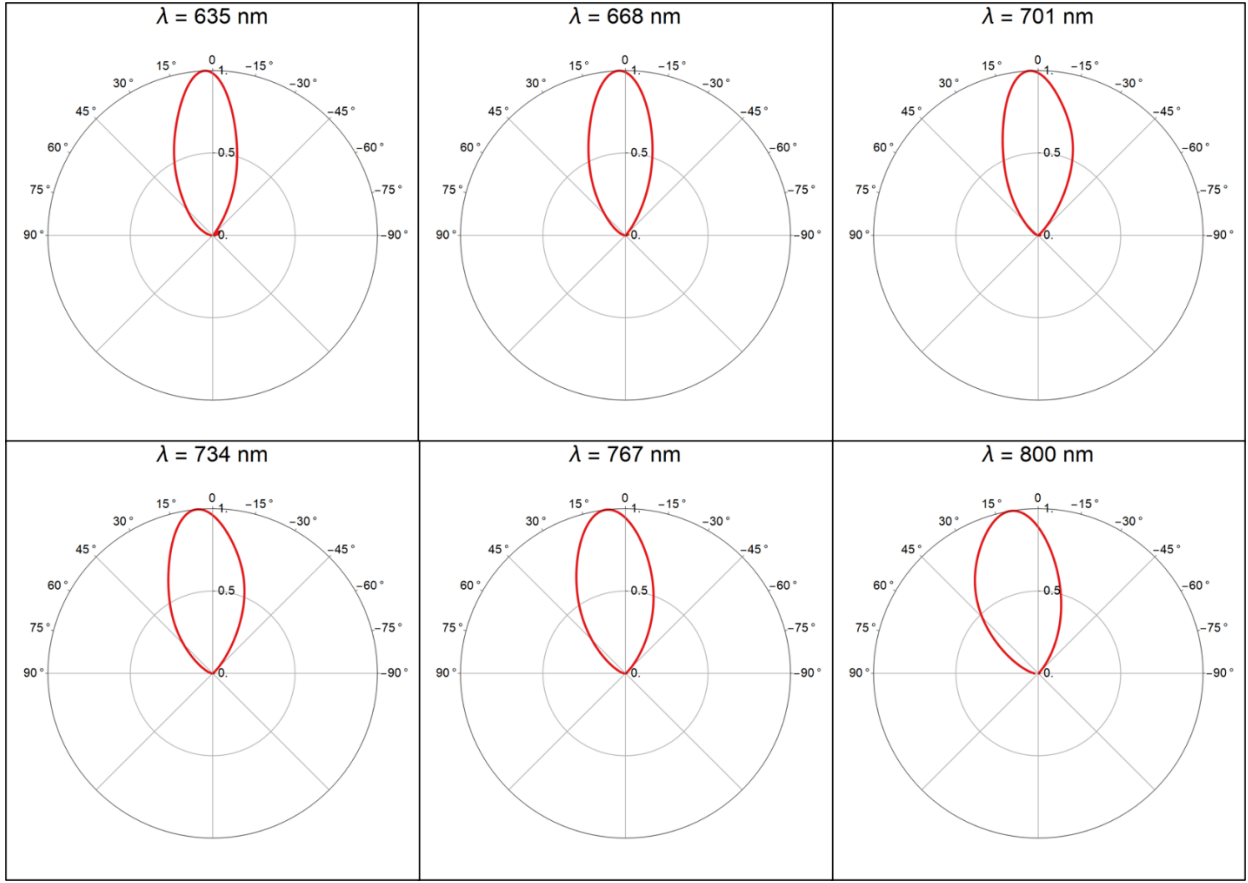

**Figure S3.** To make sure the beaming capabilities of our device are maintained over the full 635 – 800 nm spectrum we simulated and calculated the far-field |E|² intensities for multiple other wavelengths besides the one presented in Figure 3(d). Here we plot slices of the field profiles at an azimuthal angle of 0° and along a polar angle of ±90° for wavelengths 635, 668, 701, 734, 767, and 800 nm. The far-fields are normalized to the peak intensity over the full hemispherical projection ($-90° \leq \theta \leq 90°$ and $0 \leq \varphi \leq 360°$) so the peaks in some plots occur at another azimuthal angle. The peak intensity always stays within ±15° of the normal with only minor variations between the different wavelengths.

## S4. Collection efficiency, reflection, and absorption

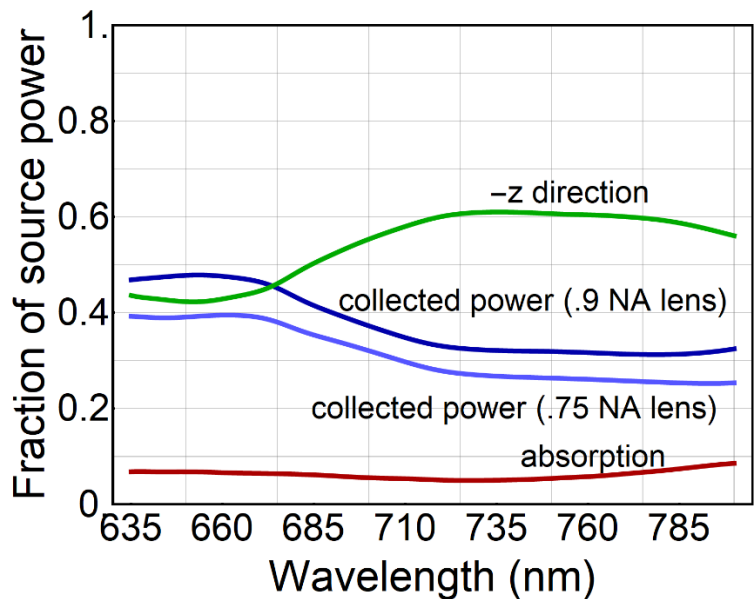

**Figure S4**. Breakdown of light emitted by the NV center in the presence of the NLE. The collection efficiency is defined as the fraction of emitted power can be collected by a lens with a specific NA. The power recorded in the -z direction is defined as the power flow downwards through a plane placed 100 nm below the NV center and diamond interface. The absorption is calculated as 1 – (power going towards +z) – (power going towards -z). The power flow is normalized to the full, Purcell-enhanced output power of the dipole source and averaged between the two dipole orientations.